\documentclass[]{spie}  

\usepackage[]{graphicx}

\title{CAST: A search for solar axions at CERN} 

\author{J.I. Collar\supit{a},
S.~Andriamonje\supit{b}, E.~Arik\supit{c},
D.~Autiero\supit{d}, F.~Avignone\supit{e},
K.~Barth\supit{d}, E.~Bingol\supit{f},
H.~Brauninger\supit{g}, R.~Brodzinski\supit{h},
J.~Carmona\supit{i}, E.~Chesi\supit{d,j},
S.~Cebrian\supit{i}, S.~Cetin\supit{c}, G.~Cipolla\supit{d},
R.~Creswick\supit{e},
T.~Dafni\supit{f}, M.~Davenport\supit{d},
R.~De~Oliveira\supit{d},
S.~Dedoussis\supit{k}, 
L.~Di~Lella\supit{d}, C.~Eleftheriadis\supit{k},
G.~Fanourakis\supit{l}, H.~Farach\supit{e}, E.~Ferrer\supit{b}, 
H.~Fischer\supit{m}, F.~Formenti\supit{d},
T.~Geralis\supit{e}, I.~Giomataris\supit{b},
S.~Gninenko\supit{n}, N.~Goloubev\supit{n},
R.~Hartmann\supit{g}, M.~Hasinoff\supit{d,j},
D.~Hoffmann\supit{f}, I. G. ~Irastorza\supit{d}, 
J.~Jacoby\supit{f},
D.~Kang\supit{m}, K.~K\"{o}nigsmann\supit{m},
R.~Kotthaus\supit{o}, M.~Krcmar\supit{p},
M.~Kuster\supit{g}, B.~Lakic\supit{p},
A.~Liolios\supit{k}, A.~Ljubicic\supit{p},
G.~Lutz\supit{o}, G.~Luzon\supit{i},
H.~Miley\supit{h}, A.~Morales\supit{i},
J.~Morales\supit{i}, M.~Mutterer\supit{f},
A.~Nicolaidis\supit{k}, A.~Ortiz\supit{i},
T.~Papaevangelou\supit{k}, A.~Placci\supit{d},
G.~Raffelt\supit{o}, H. ~Riege\supit{d},
M.~Sarsa\supit{i}, I.~Savvidis\supit{k},
R.~Schopper\supit{f}, Y.K.~Semertzidis\supit{q}, 
C.~Spano\supit{m}, M.~Stipcevic\supit{p}, V.~Vasileiou\supit{k}, J.~Villar\supit{i},
B.~Vullierme\supit{d}, L.~Walckiers\supit{d},
K.~Zachariadou\supit{l},
K.~Zioutas\supit{d,k}
\skiplinehalf
\supit{a}Enrico Fermi Institute, University of Chicago,
Chicago, IL, USA\\ 
\supit{b}DAPNIA, Centre d'Etudes de Saclay (CEA-Saclay)
Gif-Sur-Yvette, France\\ 
\supit{c}Department of Physics,
Bogazici University, Istambul, Turkey\\
\supit{d}European Organization for Nuclear Research (CERN), 
Geneve, Switzerland\\
\supit{e}Department of Physics and Astronomy, U. South
Carolina, Columbia, SC, USA\\
\supit{f}Institut f\"{u}r Kernphysik, Technische Universitat
Darmstadt, Darmstadt, Germany\\
\supit{g}Max-Planck-Institut f\"{u}r Extraterrestrische
Physik, MPG, Garching, Germany\\ 
\supit{h}Pacific
Northwest National Laboratory, Richland, WA, USA\\
\supit{i}Instituto de F\'{\i}sica Nuclear y Altas Energ\'{\i}as,
Universidad de Zaragoza, Zaragoza, Spain\\
\supit{j}Department of Physics and Astronomy, U.
of British Columbia, Vancouver, Canada\\
\supit{k}Aristotle University of
Thessalon\'{\i}ki, Thessalon\'{\i}ki, Greece\\ 
\supit{l}National
Center for Scientific Research "Demokritos" (NRCPS), Athens,
Greece\\ 
\supit{m}Albert-Ludwigs-Universit\"{a}t Freiburg,
Freiburg, Germany\\ 
\supit{n}Institute for Nuclear
Research (INR), Russian Academy of Sciences, Moscow, Russia\\
\supit{o}Max-Planck-Institut f\"{u}r Physik, Munich,
Germany\\ 
\supit{p}Ruder Boskovic Institute, Zagreb,
Croatia\\
\supit{q}Brookhaven National Laboratory, NY,
USA
}

\authorinfo{E-mail: collar@uchicago.edu}

\begin{document} 
  \maketitle 

\begin{abstract}
The new axion helioscope at CERN started acquiring data 
during September of 2002: 
CAST (Cern Axion Solar Telescope) employs a decommissioned 
LHC dipole magnet to convert 
putative solar axions or axion-like particles 
into detectable photons.
The unprecedented dipole magnet intensity and length 
(9.5 T, 10 m) results in a projected 
sensitivity that surpasses astrophysical constraints on these 
particles for the first time, 
increasing the chance of discovery. The use of X-ray focusing optics and 
state-of-the-art detector technology has led to an extremely low background 
for an experiment above ground. A brief status report is given, with emphasis on 
the tracking and control system and possible future extensions.
\end{abstract}



\section{INTRODUCTION}
\label{sect:intro}  

Axions entered the particle physics arena as a possible 
solution to the so-called strong $C\!P$ problem\cite{peccei}.
They soon became more attractive with the realization that for some 
mass ranges
they are prime galactic Dark Matter candidates  
\cite{sikivie}. To add to their
interest, if axions exist, they should be copiously produced in stellar
interiors primarily via the Primakoff effect in the scattering of thermal
photons off nuclei \cite{jorge,rosenberg}. 
In the case of our Sun, theoretical expectations are
for a low-energy axion emission spectrum peaked around a mean energy of
$\sim$4.4 keV and dying off at $\sim$10 keV, 
while in supernova explosions they could carry up to
a much larger $\sim$160 MeV \cite{jorge,rosenberg}. 
Axions can participate in
stellar energy dissipation mechanisms, affecting stellar
evolution to the point that useful (but somewhat uncertain)
theoretical limits have been obtained from, for instance, the life span of
Horizontal Branch (HB) stars \cite{jorge,rosenberg}.
The Sun is, due to its closeness to the Earth, the astrophysical source 
of choice for axion
searches. 
The solar neutrino problem added a humbling dash of
uncertainty to our knowledge of a star's inner mechanisms: more to the 
point,
the role that yet undiscovered particles might 
play, be those axions
or axion-like (i.e., participating of some axion
couplings) is still the subject of great theoretical and experimental interest. 
In this sense it is important to emphasize that any new 
scalar particle can couple to two photons via fermion 
(quark and lepton) vacuum loops \cite{smith} (\frenchspacing{Fig. 1}). 
Experiments exploiting the Primakoff mechanism are therefore 
capable of discovering any of a large number of proposed particles arising 
from broken or unbroken symmetries (Majoron, 
Familon, Paraphoton, and so on).
   \begin{figure}
   \begin{center}
   \begin{tabular}{c}
   \includegraphics[height=3cm]{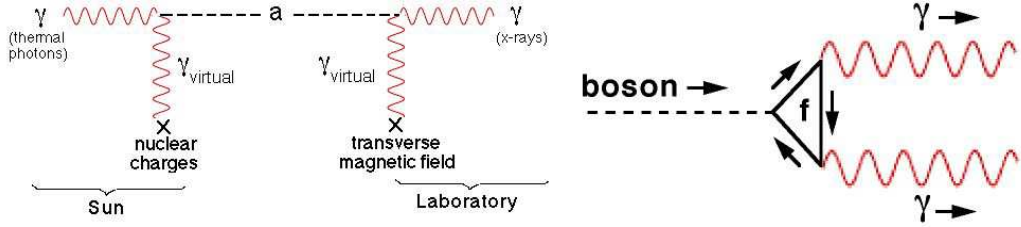}
   \end{tabular}
   \end{center}
   \caption[fig1] 
   { \label{fig:fig1} 
Left: The Primakoff effect (the coupling of an axion to 
two photons) at work in the Solar core and in the 
laboratory. Right: A new scalar particle coupling to charged particles is
subject to this effect. 
For this reason, experiments such as CAST are 
generic searches, not limited 
to Peccei-Quinn invisible axions in their discovery potential.}
   \end{figure} 
   
\section{CERN AXION SOLAR TELESCOPE (CAST)} 
The detailed conceptual design of a solar axion telescope, including a
description of previous experiments, 
calculations of expected sensitivity, detector background estimates, \frenchspacing {etc.},
is described in \cite{bibber,us}.
In essence, the working principle is as follows: 
An incoming axion couples to a virtual photon provided by the transverse
field ($B$) of an intense dipole magnet, 
being transformed into a real, detectable photon that
carries the energy and momentum of the original axion
($axion + \gamma_{virtual} \rightarrow \gamma$, \frenchspacing{Fig. 1}). 
Hence, the magnetic field $B$ plays the role of a catalyst. 
Low-background X-ray detectors at the far end of the magnet are 
sensitive to the conversion photons, yet exclusively at times of alignment 
between the magnet and the Sun, providing a unique axion signature. 

For the axion energies and rest masses of concern in
such searches,  the above interaction is coherent, i.e., the axion-to-photon
conversion probability is proportional to $(B\cdot L)^2$, where $L$ is the
active length of the magnet bore\cite{us}. 
A decommissioned straight-bore
LHC test magnet of $B\sim9.6$ T, $L\sim9.5$ m and
$\sim 10$ mrad angular opening has provided a rare opportunity for 
the construction of a high-sensitivity axion telescope, the CAST 
experiment at CERN \cite{us,wwwCAST} (\frenchspacing{Fig. 2}).
A single one of these magnets ($B\cdot L=91~T\cdot m$) 
is $\sim 100$ times more efficient as an axion-to-photon
converter than the best competing setup, presently in operation at the
University of Tokyo ($B\cdot L=9.2~T\cdot m$)
\cite{japan}. The achievable Primakoff coupling sensitivity for light axion 
masses ($< 1$ eV) is
approximately given by:
\begin{equation}
g_{a\gamma\gamma}\leq 1.4\cdot 10^{-9} [\textrm{GeV}^{-1}]
~\frac{b^{1/8}}{t^{1/8}~B[\textrm{T}]^{1/2}~L[\textrm{m}]^{1/2}~A^{1/4}}~,
\end{equation}
where $b$ is the detector background in counts/day in the
energy region $\sim 1-10$ keV, $t$ [days] is the time of 
alignment of the magnet bore with  the Sun and $A [cm^{2}]$ is the bore 
opening area (14 $cm^{2}$ for this magnet). In view of the 
weak dependence on all factors other than $B$ and $L$, it will be 
extremely hard
to improve the axion 
sensitivity of CAST with present or foreseeable magnet technology.
For the first time and over many orders of magnitude 
in axion mass, the experimental sensitivity will surpass 
the astrophysical axion constraints (\frenchspacing{Fig. 3}).
   \begin{figure}
   \begin{center}
   \begin{tabular}{c}
   \includegraphics[height=6cm]{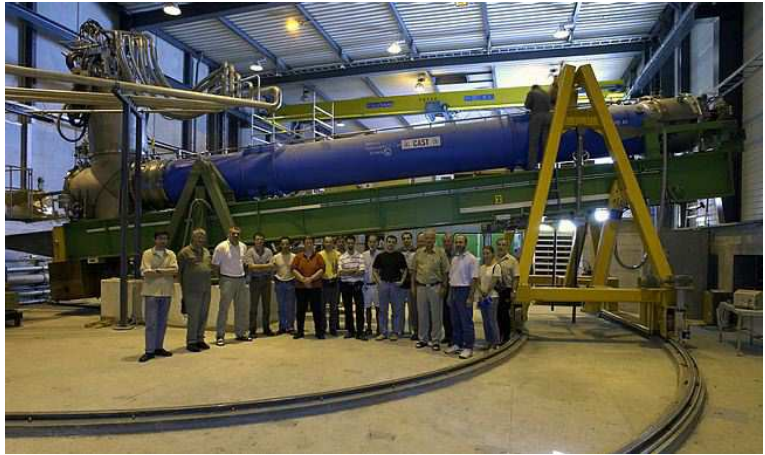}
   \end{tabular}
   \end{center}
   \caption[fig2] 
   { \label{fig:fig2} 
Members of the CAST collaboration in front of the helioscope during 
the first week of data acquisition (September 2002). Visible to the 
left are the flexible cryogenic pipelines that allow a free magnet 
motion. The TPC is visible on the right end.}
   \end{figure} 
   \begin{figure}
   \begin{center}
   \begin{tabular}{c}
   \includegraphics[height=7cm]{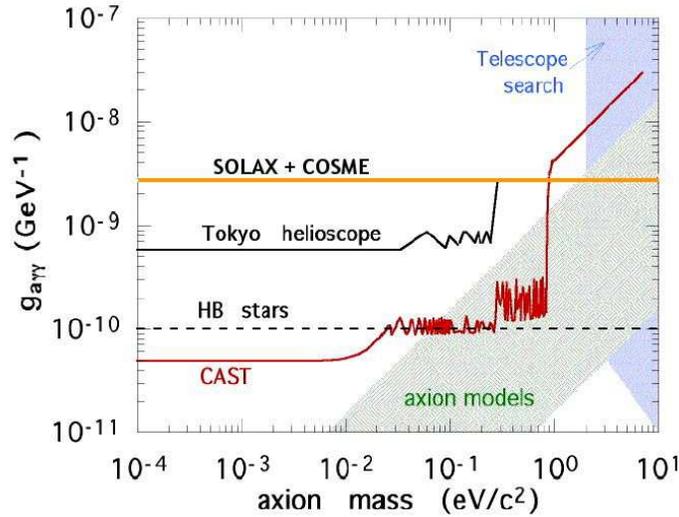}
   \end{tabular}
   \end{center}
   \caption[fig1] 
   { \label{fig:fig3} 
Expected sensitivity for CAST.
Also shown: limits obtained using low-background germanium detectors
(SOLAX, COSME \protect\cite{cosme}), those from the Tokyo
helioscope \protect\cite{japan}, the
theoretical red giant bound\protect\cite{red_giant}, the 
region excluded from the absence of an axion-decay
photon line in galactic
clusters\protect\cite{indirect}, and the region favored by theoretical axion
models. A further increase in CAST's sensitivity by a factor $\sim$2 seems 
feasible from the combined use of X-ray focusing optics and 
micropattern detectors (see text).}
\end{figure} 

Three different detectors have been installed,
all of them sensitive to the X-rays that might originate in the conversion of axions
inside the magnet: a time projection chamber (TPC), a CCD and a
MICROMEGAS chamber \cite{mms}. The TPC exhibits the
robustness of a well-tested technique, even if its spatial resolution
(and consequently background rejection capability)
is inferior to the other two. It is constructed of selected 
low-background materials and covers both magnet bores sensitive 
to "sunset"
axions. Facing "sunrise" axions,
the CCD detector and MICROMEGAS chamber
work in conjunction with a mirror system able to focus any X-rays
streaming out of the magnet bores. This system is yet another example of 
recycling, as it employs an engineering unit left behind by the ABRIXAS 
X-ray astronomy
mission. The mirror concentrates the sought signal into a $\sim$1 
mm$^{2}$ spot. 
This allows to use a small detector (or one with high spatial resolution) 
therefore increasing the signal-to-background ratio
by roughly two orders of magnitude. The expectations in these 
conditions are essentially those of a ``zero-background'' experiment: 
in the absence of axions, only
$\sim$1 count/month is expected after energy and spatial cuts.
The detectors and mirror are
already installed on their corresponding ends of the magnet. At the time of
this writing, several hours of data have been
taken with the TPC in "axion-sensitive" conditions, i.~e., with
the magnet on (9.4 T) and active tracking of the Sun. Routine 
operation with all detectors online is expected to start in February of 
2003, spanning a period of three years under different running 
conditions (with vacuum or He gas in the bores \cite{us}).

The solar tracking system has been checked using
a small alignment telescope mounted on top of the 
magnet (and carving windows on the otherwise solid walls of the 
experimental hall). Footage from these satisfactory tests  
is available at 
\cite{pelis}. 
The first step in this critical part of the project was to measure the 
orientation of the magnet in the relevant
set of topocentric coordinates (azimuthal angle AZ and zenith distance 
ZD) for ninety platform positions, creating a look-up 
table of AZ and ZD values correlated to position encoder readings. 
The collaboration of the EST division at CERN  
has proven invaluable for this.
These measurements will be repeated along the duration of the project 
to account for possible settling or flexure of the structure. They 
are performed with a precision better than $0.001^{\circ}$. As a 
reference, the angle subtended by the magnet bore is $\sim 0.25^{\circ}$, 
while the solar core spans $\sim 0.05^{\circ}$ (the bulk of axion 
emission is expected from this inner $\sim$1/10 of the solar radius). The second step
is to generate reliable solar AZ and ZD 
predictions with better than $0.01^{\circ}$ accuracy \cite{novas} and to direct 
the platform (via control of two motors) to the corresponding encoder values using a
non-cartesian spline interpolation of the 
look-up tables (``Hardy's multiquadrics'' 
\cite{spline}). Finally, all components are merged into a stand-alone LABVIEW
application also able  to log all environmental parameters of 
interest. The precision goal ($<0.01^{\circ}$ error) 
of the resulting fully-automatic tracking 
system has been met 
in both topocentric and galactic coordinates (\frenchspacing{Fig. 4}).
The evolution of the experiment
can be followed live (with hourly updates) \cite{wwwLV}. 
   \begin{figure}
   \begin{center}
   \begin{tabular}{c}
   \includegraphics[height=6cm]{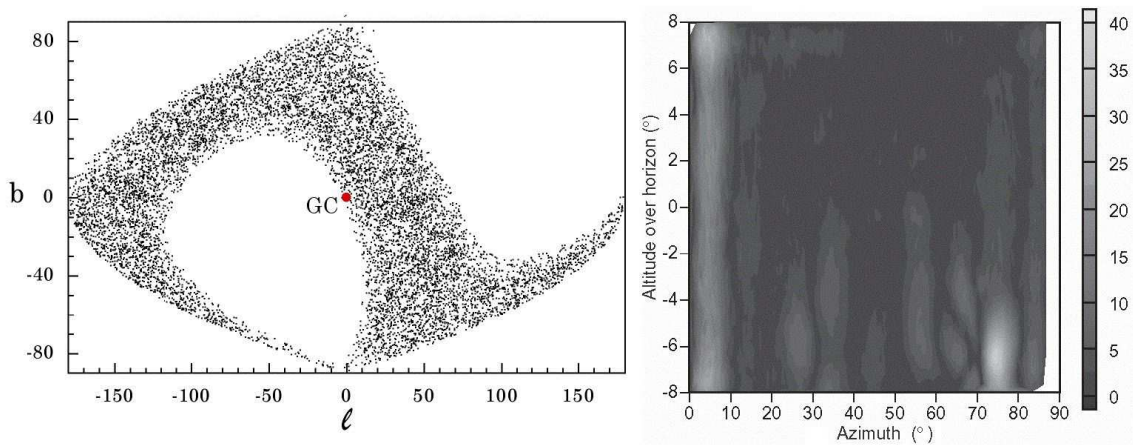}
   \end{tabular}
   \end{center}
   \caption[fig4] 
   { \label{fig:fig4} {\it Left:} Patch of sky scannable with CAST's 
final range of motion ($\pm40^{\circ}$ horizontal, $\pm 8^{\circ}$
vertical), in galactic 
coordinates. The automatic platform control system is 
able to track any  user-defined reachable set of galactic coordinates during 
the hours when 
solar tracking is impossible. Solar tracking  
can be followed live at \protect\cite{wwwLV} (webpage updated every hour)
{\it Right:} Maximum tracking precision attainable by 
the platform control system as a function of spatial direction. The 
grey-scale units are thousands of a degree. The azimuth origin in the 
figure is 
offset to the NE direction.
}
\end{figure} 

An interesting possible future direction is the development of low-background detectors
able to perform parasitic searches for cosmological 
higher-energy 
axion-like particles. While a mapping of the sky in 
galactic coordinates looking 
for an excess signal along the galactic plane or center
might be of interest {\it per se} ( ``axion astronomy''?), a more realistic high-energy boson 
search should continue to rely on the Sun 
as a source: If a new boson couples to nucleons, it can 
substitute for a photon in a number of solar plasma and nuclear processes 
\cite{jorge}. Weak experimental limits already exist from the observed flux of
solar gamma-rays below 5.5 MeV, to which axion decay 
($a\rightarrow\gamma\gamma$) following $p + d \rightarrow ^{3}\!\!\!He + a$ 
might contribute\cite{jorgeandleo}. Other unexplored 
interesting channels exist\cite{jorge,jorgeandleo,ketal}  
and a generic search should not be limited to pseudoscalar particles (i.e., 
M1 transitions like the one mentioned). This question has not been 
examined in all its generality, leaving room for surprises\cite{jorge2}. 
It must be kept in mind that astrophysical constrains
still allow axions to 
partake of a non-negligible fraction 
(few \%)
of the total solar luminosity \cite{jorge}.
In particular, the absence of a 511 keV 
excess signal (from $e^{+} + e^{-} 
\rightarrow \gamma + a$) in these additional CAST detectors
at times of solar alignment may impose tighter 
bounds than similar searches for anomalous production of single 
photons in accelerator experiments \cite{511}. 
Other ideas are presently 
under study (e.g., study of solar emission of light pseudoscalars in the visible or UV, 
etc.).

\bibliography{report}   
\bibliographystyle{spiebib}   

\end{document}